\documentclass[aps,prl,a4paper,twocolumn,showpacs,floatfix,citeautoscript,superscriptaddress,footinbib]{revtex4-1}
\usepackage{graphicx}
\usepackage{amsfonts}
\usepackage{amsmath}
\usepackage{amssymb}
\usepackage{bm}

\usepackage{bbold}
\usepackage{mathtools}
\usepackage{dcolumn}
\usepackage{hyperref}
\usepackage[usenames]{color}
\usepackage{subcaption}

\usepackage{physics}
\usepackage{units}

\usepackage{ragged2e}
\DeclareCaptionJustification{justified}{\justifying}
\newcommand{\out}[1]{{}}

\def\vk{{\bf k}}

\begin{document}

\title{
Correlated electronic structure and optical response of rare-earth-based semiconductors 
}
\author{Anna Galler}
\address{Institute of Solid State Physics, TU Wien, 1040 Vienna, Austria}
\address{Centre de Physique Th\'eorique, Ecole Polytechnique, CNRS, Institut Polytechnique de Paris,
  91128 Palaiseau Cedex, France}
\author{James Boust}
\address{Centre de Physique Th\'eorique, Ecole Polytechnique, CNRS, Institut Polytechnique de Paris,
  91128 Palaiseau Cedex, France}
\author{Alain Demourgues}
\address{Institut de Chimie de la Matiere Condens\'ee de Bordeaux (ICMCB), 33600 Pessac, France}

\author{Silke Biermann}
\address{Centre de Physique Th\'eorique, Ecole Polytechnique, CNRS, Institut Polytechnique de Paris,
  91128 Palaiseau Cedex, France}
\address{Coll\`ege de France, 11 place Marcelin Berthelot, 75005 Paris, France}
\address{Department of Physics, Division of Mathematical Physics, Lund University, Professorsgatan 1, 22363 Lund, Sweden}
\address{European Theoretical Spectroscopy Facility, 91128 Palaiseau, France, Europe}

\author{Leonid V. Pourovskii}
\address{Centre de Physique Th\'eorique, Ecole Polytechnique, CNRS, Institut Polytechnique de Paris,
  91128 Palaiseau Cedex, France}
\address{Coll\`ege de France, 11 place Marcelin Berthelot, 75005 Paris, France}

\begin{abstract}
The coexistence of Mott localized $f$-states with wide conduction and valence bands in $f$-electron
semiconductors results, quite generically, in a complex optical  response with the nature of absorption edge difficult to resolve both experimentally and theoretically.
Here, we combine a dynamical mean-field theory approach to localized 4$f$ shells with an improved description of band gaps by a semi-local exchange-correlation potential  
to calculate the 
optical properties of the light rare-earth fluorosulfides $Ln$SF ($Ln=$Pr, Nd, Sm, Gd) from first principles.  
 In agreement with experiment, we find the absorption edge in SmSF  to stem from S-3$p$ to Sm-4$f$ transitions, while the Gd compound  behaves as an ordinary  $p-d$ gap semiconductor. In  the unexplored PrSF and NdSF systems we predict  a rather unique occurrence of strongly hybridized 4$f$-5$d$ states at the bottom of the conduction band.  The nature of the absorption edge results in a characteristic anisotropy of the optical conductivity in each system, which may be used as a fingerprint of the relative energetic positions of different states.
\end{abstract}

\maketitle

The electronic structure of rare-earth based semiconductors -- rare-earth oxides, sulfides, chalcogenides, etc. -- is determined by the coexistence of wide, ligand $p$ and rare-earth 5$d$,  semiconducting bands with localized rare-earth 4$f$ states.  The latter form characteristic sharp quasi-atomic peaks in the electronic spectra and, when occurring inside the semiconducting $p$-$d$ gap, determine the onset of optical absorption. The resulting sharp absorption edges render rare-earth semiconductors promising as 
less-toxic and environmentally-benign pigment materials~\cite{ce_yellow,ti_yellow,la_yellow_orange,re_blue,masui_green_cuprate}.
The nature of the optical transitions is often quite challenging to resolve in these materials, even experimentally, since the position and structure of the quasi-atomic 4$f$ bands vary significantly along the series.

The localization of  the 4$f$ electrons is a Mott phenomenon induced by a large on-site Coulomb repulsion; a simultaneous quantitative description of both the  $p$-$d$ band gap and 4$f$ Mott gap is then a challenge even for a qualitative theoretical description.  The quasi-atomic rare-earth 4$f$ shells harbor strong local correlation effects, while the formation of the $p$-$d$ semiconducting gap between ligand $p$ and rare-earth 5$d$ bands involves non-local exchange and correlation. Both phenomena lead to spectral signatures that cannot be comprehensively addressed within standard density functional approaches (DFT) and call for many-body techniques such as the dynamical mean-field theory (DMFT)~\cite{Vollhardt_DMFT,Georges_DMFT} and many-body perturbation theory (G$W$)~\cite{Hedin1965,Hedin1999,Biermann_GW+DMFT}. 

Among  rare-earth based semiconductors, several  families of chalcogenides -- $Ln$SF, oxysulfides $Ln_2$O$_2$S and oxyfluorosulfides $Ln_3$OF$_3$S$_2$ (with $Ln$=Ce, Pr, Nd, Sm, Gd) -- feature ionic fluorine or oxygen 2$p$ bands together with more covalent sulfur 3$p$  bands and localized rare-earth 4$f$ states. Some of these materials exhibit  brilliant colors ranging from yellow (SmSF, GdSF) to red (CeSF) and have recently attracted attention as a possible replacement for the common, but more toxic, red-yellow pigments based on cadmium. Intensive experimental studies of the $Ln$SF series by optical spectroscopy and X-ray photoemission~\cite{demourgues_lnsf_pigments,demourgues_preparation,pauwels_structural_features,pauwels_thesis} found significant modifications of the shape and position of the 4$f$ states, concomitant with a shift of the absorption edge to smaller wavelengths from CeSF (597 nm) to SmSF and GdSF (490 nm) along the series. This shift of the absorption edge has been related to the evolution of the ionization energies along the rare-earth series \cite{demourgues_lnsf_pigments}.   This analysis suggests that the nature of the absorption edge changes along the series, e.g. from 3$p \to$4$f$ in SmSF to  3$p \to$5$d$ in GdSF. The optical properties of two other light rare-earth $Ln$SF -- PrSF and NdSF -- have not been measured so far.
Initial theoretical calculations~\cite{Goubin2004} attempted to evaluate the electronic spectra and dielectric function of $Ln$SF using standard DFT, which faces difficulties in treating the local 4$f$ correlations and  $p$-$d$ semiconducting gap correctly.
A  detailed theoretical study combining DFT and DMFT has been carried out only for the red pigment CeSF~\cite{pnas_cesf}, without however attempting to determine the $p$-$d$  gap within a first-principles framework.

In this work we use an \textit{ab initio} methododology to calculate the electronic structure and optical properties of the light  rare-earth fluorosulfides $Ln$SF. The very subtle task of identifying the optical transitions in $Ln$SF from first principles is addressed by taking into account both local correlation effects and non-local exchange-correlation effects; this is crucial for determining the relative positions of the S-3$p$ states, the rare-earth 5$d$ states and the correlated rare-earth 4$f$ states.
We focus on systems with localized 4$f$ shell  ($Ln=$Pr, Nd, Sm, Gd), where  one can confidently apply  a quasi-atomic approximation to the 4$f$ local correlations within DMFT. An improved treatment of the semiconducting $p$-$d$ gap is provided by a non-local exchange-correlation potential. 
Our calculated many-body spectral functions and anisotropic optical conductivities show a non-trivial evolution of the optical absorption edge along the series. In particular, we predict an absorption edge formed by optical transitions from the S-3$p$ to a strongly hybridized rare-earth 5$d$-4$f$ band in PrSF and NdSF. We also confirm the different nature of the absorption edge in SmSF and GdSF, as previously proposed in Ref.~\cite{demourgues_lnsf_pigments}. In each compound our calculations identify a particular anisotropy of the optical conductivity, which, as we show, represents a fingerprint of the states involved in the absorption-edge transitions. 

Our methodology combines
electronic structure~\cite{Wien2k} and many-body methods by treating the correlated rare-earth 4$f$ shell within DMFT~\cite{Anisimov1997_1,Kotliar_review,Aichhorn2009,Aichhorn2011,Pourovskii2007,triqs_cpc_2015,triqs_dfttools,Beach2004}, and employing the modified Becke-Johnson~\cite{tran_mbj_original,koller_mbj_2011,hong_mbj_2013} exchange-correlation potential in DFT  to take into account the effect of non-local exchange onto the $p$-$d$ band-gap.
Spin-orbit coupling is included. On-site correlation effects between 4$f$ electrons are evaluated within a quasi-atomic (Hubbard-I~\cite{hubbard_1}) approximation using a fully rotationally-invariant Coulomb interaction. We abbreviate this approach as mBJ+Hubbard-I.  Cross-checks with a numerically exact  continuous-time quantum Monte Carlo method~\cite{Gull2011,ctqmc_werner} (see Supplementary) reveal only small \unit[0.1-0.15]{eV} upward shifts of the unoccupied 4$f$ states due to an underestimation of hybridization effects in mBJ+Hubbard-I. The corresponding  correction for hybridization effects, extracted as described in Supplementary, is subsequently included into our mBJ+Hubbard-I framework.  
The  on-site screened Coulomb interaction between rare-earth 4$f$ electrons is calculated from first principles using a combined constrained local density approximation (cLDA)+Hubbard-I approach~\cite{galler_cLDA} that treats the quasi-atomic rare-earth 4$f$ states appropriately. 
For details regarding our methodology we refer the reader to the Supplementary material and Ref.~\cite{James_paper}. 
As the result of our calculations, we obtain the many-body spectral function $A_k(\omega)$ encoding the excitation energies of an electron addition/removal into the many-body ground state as well as optical response functions (optical conductivity).  
 
We calculate the rare-earth fluorosulfides $Ln$SF ($Ln=$ Pr, Nd, Sm, Gd) in their experimental  tetragonal PbFCl-type crystal structure (space group $P4/nmm$). The lattice parameters $a$ and $c$ decrease continuously from PrSF to GdSF due to the reduction of the $Ln$ ionic size along the rare-earth series (see Ref.~\cite{demourgues_lnsf_pigments} for the lattice parameters and a detailed description of the $Ln$SF crystal structure).

\begin{figure}[]
\includegraphics[width=0.9\columnwidth]{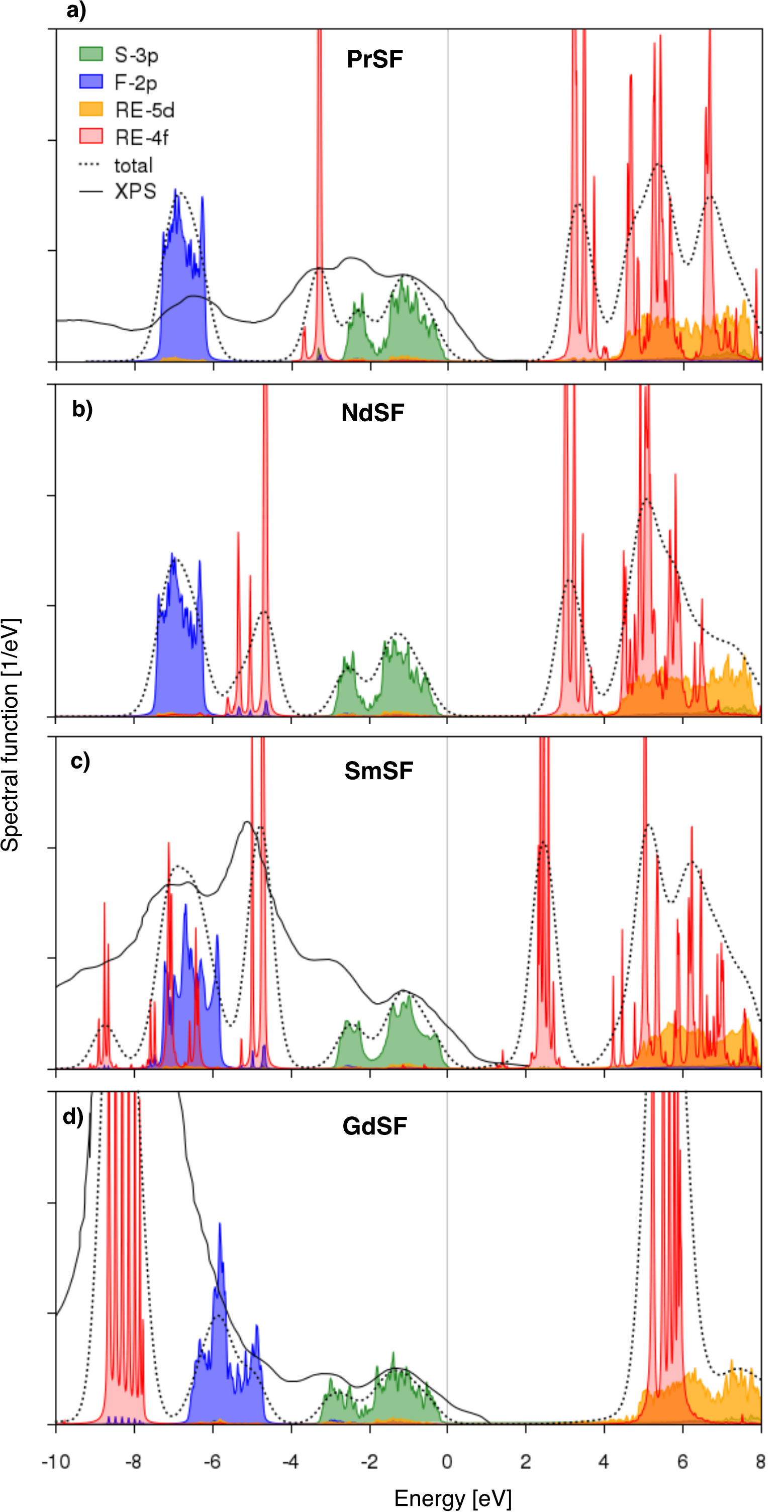}
\caption{\textbf{Many-body spectral functions compared to experimental XPS spectra.} While the F-2$p$ (blue), the S-3$p$ (green) and the rare-earth 5$d$ (orange) spectral weight  remain almost unchanged from PrSF to GdSF \textbf{(a)-(d)}, the rare-earth 4$f$ states show a significant evolution which defines the absorption-edge  optical transitions. The position of the peaks in the total broadened spectral function of the occupied states (dotted black lines, broadening=\unit[0.4]{eV}) compares well to the experimental XPS spectra (solid black lines, taken from Ref.\cite{pauwels_thesis}.) }
\label{fig:dos}
\end{figure}

\begin{figure*}[t!]
	\begin{minipage}{18 cm}
		\includegraphics[width=18.cm]{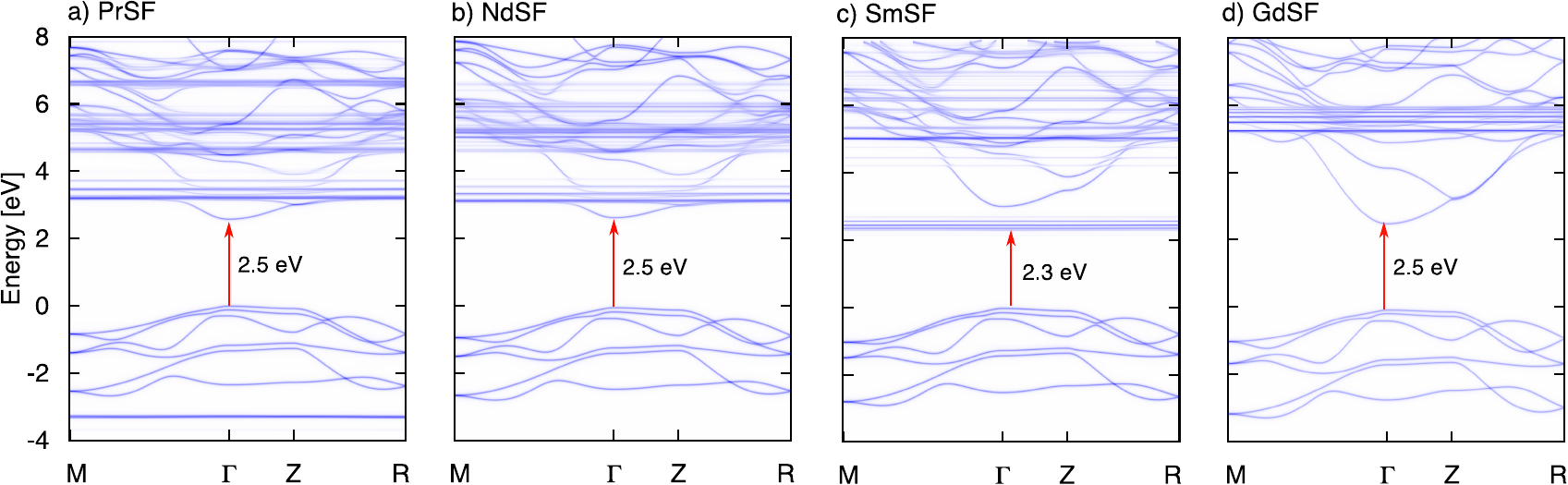}
		\caption{\textbf{$\vk$-resolved spectral functions of the rare-earth fluorosulfides $Ln$SF.} \textbf{(a)} PrSF and \textbf{(b)} NdSF show an optical gap of \unit[2.5]{eV} formed between the S-3$p$ band (highest occupied bands) and a strongly hybridized rare-earth 5$d$-4$f$ band. \textbf{(c)} In SmSF, the optical absorption edge can clearly be attributed to transitions from S-3$p$ to Sm-4$f$ (flat bands around \unit[2.3]{eV}). \textbf{(d)} In GdSF, the Gd-4$f$ states are not involved in the optical transitions and the optical gap is a pure S-3$p$ to Gd-5$d$ band-gap.  
			\label{fig:bands} }
	\end{minipage}
\end{figure*}

Let us first focus on the spectral functions of $Ln$SF.  The $\vk$-integrated spectral functions for $Ln$SF, calculated with our \textit{ab initio} methodology, are shown in Fig.\ref{fig:dos} and compared to experimental X-ray photoemission (XPS) spectra from Ref.~\cite{pauwels_thesis}.   
All $Ln$SF compounds share similar features related to the F-2$p$,  S-3$p$ and $Ln$-5$d$ states.
The spectral weight at around \unit[-6]{eV}  stems from the F-2$p$ states (we measure all energies with respect to the top of the S-3$p$ band). Due to the high electronegativity of fluorine, 
the F-2$p$ states are located, as expected, well below the S-3$p$ band. Notice, however, a significant reduction of the F-2$p$ binding energy in the case of GdSF. This effect is due to an evolution of the bonding length along the series, with the single apical $Ln$-S bond in GdSF becoming  shorter than the four planar $Ln$-S bonds. The resulting charge redistribution apparently weakens the ionic Gd-F bond.
In the unoccupied part of the spectral function one identifies the $Ln$-5$d$ states separated from S-3$p$ by the band gap. 

Although the F-2$p$, the S-3$p$ and rare-earth 5$d$ states remain almost unchanged throughout the $Ln$SF series, the rare-earth 4$f$ states exhibit a pronounced evolution. Due to their localized character, the 4$f$ states form sharp peaks, corresponding to transitions between quasi-atomic multiplets, and hybridize only weakly with the rest of the states.  In PrSF, the  4$f$ states are split into an occupied lower Hubbard band situated slightly above \unit[-4]{eV}, and an unoccupied upper Hubbard band above \unit[2.5]{eV}. The upper Hubbard band in the Pr, Nd, and Sm compounds is split into separate peaks due to multiplet effects, with the total upper Hubbard band width of about \unit[5-6]{eV}; these effects are absent for the half-filled Gd-4$f$ shell resulting in a much more narrow upper Hubbard band.
  Examining the position of the occupied 4$f$  states along the series in  Fig.~\ref{fig:dos}, one notices their progressive shift to lower energies, from PrSF to GdSF. This evolution is  due to the well-known increase of the 4$f$ binding energy along the rare-earth series. 

The separation between the 4$f$ lower and upper Hubbard band is due to the on-site Coulomb repulsion in the rare-earth 4$f$ shell. The screened Coulomb interaction $U$ for the 4$f$ shell has been calculated using our \textit{ab initio} cLDA+Hubbard-I method. The corresponding values are given in Supplementary Table I, together with the experimental Hund's coupling $J_H$ extracted from Ref.~\onlinecite{carnall_lanthanides_1989}; $J_H$ for the rare-earth 4$f$ states is known not to be sensitive to the crystalline environment. 
For PrSF, NdSF and SmSF the distance between the lower and upper Hubbard band is approximately given by $U-J_H$, while it is $U+6J_H$ for GdSF. This is due to the half-filled Gd 4$f$ shell having the maximum value of spin (Hund's rules) and thus being particularly stable.  
In Fig.~\ref{fig:dos}d) one can see that the 4$f$ lower Hubbard band in GdSF is located at around \unit[-8]{eV}, while the upper Hubbard band is above \unit[5]{eV}. Hence, for GdSF the 4$f$ states are not involved in the optical transitions forming the optical absorption edge of \unit[2.5]{eV}. For the other three investigated compounds the unoccupied 4$f$ states are lying lower and play an important role in determining the optical gap.

\begin{figure*}[t!]
	\begin{minipage}{18 cm}
		\includegraphics[width=18.cm]{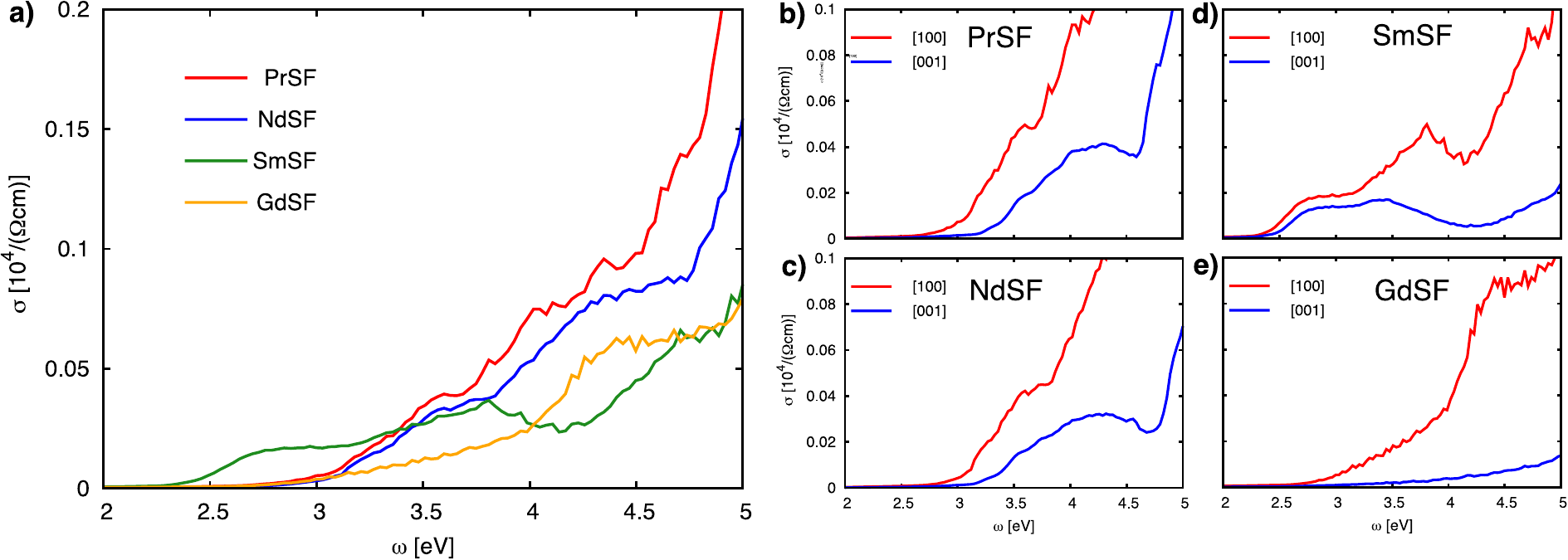}
		\caption{ \label{fig:opt_cond} \textbf{Anisotropic optical conductivities.} \textbf{(a)} Comparison of the polarization-averaged optical conductivities.  \textbf{(b)-(e)} The conductivities  for two different orientations of the light polarization, [100] (in red) and [001] (in blue). The strong anisotropy allows to distinguish between $p$-$f$ (in SmSF) and $p$-$d$ (in GdSF) absorption edges. In PrSF and NdSF one has a $p$-$d$ absorption edge, which is immediately followed by $p$-$f$ transitions above \unit[3]{eV}.}
	\end{minipage}
\end{figure*}

In Fig.~\ref{fig:dos} we further compare our calculated
spectral functions to the experimental X-ray photoemission (XPS) spectra from Ref.~\cite{pauwels_thesis}.
In order to simulate experimental resolution, the theoretical spectral functions have been convoluted
with a Gaussian with the full width at half maximum of \unit[0.4]{eV}. To approximately account for the difference in photoemission cross-section between various states---in particular the increase of relative 4$f$ cross section from PrSF to GdSF---the 4$f$ contribution has been rescaled for each compound so that the ratio between 4$f$ and 3$p$ spectral weight agrees with experiment. We observe a very good agreement of peak positions in the theoretical and experimental spectra of PrSF, SmSF and GdSF (no XPS data is available for NdSF). The S-3$p$ double-peak structure is clearly seen in both theory and experiment. Also the positions of the occupied 4$f$ and the F-2$p$ states are correctly predicted by our \textit{ab initio} calculations. In case of GdSF, the 4$f$ cross section is so large that the F-2$p$ spectral weight is absorbed into the intense Gd-4$f$ peak.
The good agreement between theory and experiment confirms the accuracy of our \textit{ab initio} calculations and their capability of predicting the nature of the optical transitions in $Ln$SF.

The unoccupied part of the spectra can more clearly be analyzed from the $\vk$-resolved spectral functions shown in Fig.~\ref{fig:bands}.  
The $\vk$-resolved spectral function of PrSF in Fig.~\ref{fig:bands}a) shows a Pr-5$d$ band in the unoccupied part of the spectrum crossing and hybridizing with the lowest unoccupied 4$f$ states, i.e. the flat bands at around \unit[3.2]{eV}. At the $\Gamma$-point, this strongly hybridized $d$ band is the lowest-lying unoccupied band defining an optical gap of \unit[2.5]{eV}. Thus, the optical absorption edge in PrSF of \unit[2.5]{eV} is formed by optical transitions from the S-3$p$ states, i.e. the highest occupied bands, to this strongly hybridized Pr 5$d$ band. These transitions are immediately followed by S-$3p$ to Pr-4$f$ transitions at around \unit[3.2]{eV}. 
As can be seen from Fig.~\ref{fig:bands}b), the situation for NdSF is very similar. Also in NdSF the optical gap of \unit[2.5]{eV} is formed between the S-3$p$ bands and a strongly hybridized band of Nd-5$d$ and Nd-4$f$ character. The lowest unoccupied 4$f$ states in NdSF are lying at \unit[3.1]{eV}, slightly lower than in PrSF.

The picture clearly changes for SmSF, shown in Fig.~\ref{fig:bands}c), where the Sm-4$f$ states are located within the band-gap of \unit[2.8]{eV} formed by the S-3$p$ and the Sm-5$d$ bands. Here, the optical absorption edge is formed by transitions from the occupied S-3$p$ states to the lowest unoccupied Sm-4$f$ states lying at \unit[2.3]{eV}. Note that the $\vk$-integrated spectral function of SmSF in Fig.~\ref{fig:dos}c shows a very small 4$f$ peak below \unit[2]{eV}; however, the intensity of this peak is so low that it does not show up in the plot of the $\vk$-resolved spectral function and does not contribute to the optical conductivity.  The optical gap of \unit[2.3]{eV} predicted by our \textit{ab initio} calculations is slightly underestimated compared to its experimental value. In experiment, SmSF shows a yellow color with an optical absorption edge of  \unit[2.4-2.5]{eV}~\cite{demourgues_lnsf_pigments}. 

As already noted above, in GdSF  (Fig.~\ref{fig:bands}d) the optical gap is of a pure $p$-$d$ character; the calculated optical gap of \unit[2.5]{eV}  agrees with the experimental absorption edge of GdSF (490 nm) and its yellow color\cite{demourgues_lnsf_pigments}.  
This confirms the quantitative accuracy of our mBJ+Hubbard-I treatment for semiconducting $p$-$d$ gaps.  

Let us now turn to the explicit calculation of the optical conductivity, i.e. the linear-response function of a material to incident light. As usual, one can limit the calculation to direct optical transitions, without any momentum transfer~\cite{Dresselhaus_opt}.  If we further neglect excitonic effects \footnote{Excitonic effects in $Ln$SF are expected to be small, since characteristic excitonic  features are absent from the measured dielectric function of LaSF and CeSF\cite{Goubin2004}. See also the relevant discussion in Supplementary of Ref.~\cite{pnas_cesf}.}, the real part of the optical conductivity reads   
\begin{align}
\label{eq:cond}
\sigma_{\alpha\alpha}(\Omega)  &=\frac{2\pi e^2 \hbar}{V}  \sum_{k}  \int d\omega \;\frac{f(\omega-\Omega/2)-f(\omega+\Omega/2)}{\Omega} \nonumber \\
&\times \text{Tr}\left\{\mathbf{A}_{k}(\omega+\Omega/2)\mathbf{v}_{k,\alpha}\mathbf{A}_{k}(\omega-\Omega/2)\mathbf{v}_{k,\alpha}\right\} , \; 
\end{align}
where $V$ is the unit-cell volume, $\Omega$ the frequency of the incident light and $f(\omega\pm\Omega/2)$ Fermi functions which ensure that transitions take place only between occupied and empty states. $\mathbf{v}_{k,\alpha}$ are matrix elements of the velocity operator in the direction $\alpha=x, y$ or $z$, and  $\mathbf{A}_k(\omega)$ are the $\vk$-resolved spectral function matrices obtained from mBJ+Hubbard-I, the trace of which is shown in Fig.\ref{fig:bands}. 

In Fig.~\ref{fig:opt_cond}a) we show the calculated optical conductivities of the $Ln$SF series. Along the series one notices a change in the onset and shape of the conductance according to the character of the absorption-edge optical transitions. Alike their spectral functions, the conductivities of PrSF and NdSF  look very similar, with conductance visibly setting in at \unit[2.7]{eV} and increasing significantly above \unit[3.2]{eV}.  In SmSF the onset of  optical conductance is noticeable already at \unit[2.3]{eV} because of the low-lying unoccupied Sm-4$f$ states in the $p$-$d$ band-gap.  One may notice a drastically different shape of  the conductivity in SmSF in the vicinity of the absorption edge,  as compared to NdSF and PrSF,  reflecting a different nature of the relevant optical transitions ($p$-$f$).  The conductivity in GdSF, with its pure $p$-$d$ band-gap, has yet another character with a shallow onset of conductivity at \unit[2.6]{eV} followed by a rather slow increase.  

Due to their non-cubic crystal structure, the $Ln$SF compounds feature a strong dependence of the
optical response on the orientation of the light polarization with respect to the crystallographic axis, see Figs.~\ref{fig:opt_cond}b)-e).
In PrSF and NdSF (Figs.~\ref{fig:opt_cond}b)-c), the onset of conductance is visible only for the [100] polarization, while above \unit[3]{eV} both directions, [100] and [001], contribute. This suggests that the S-3$p$ to rare-earth 5$d$ optical transitions, happening below \unit[3.2]{eV}, manifest themselves only  for the [100] polarized component of the incident light. The S-3$p$ to rare-earth 4$f$ transitions, starting above \unit[3.2]{eV}, seem instead to contribute to the optical conductivity in both directions. Accordingly, the drop around \unit[4.5]{eV} in the [001] optical conductivity of PrSF and NdSF reflects the multiplet splitting of the rare-earth 4$f$ states, i.e. the absence of 4$f$ states contributing to the conductivity in that frequency range.
This picture is confirmed when looking at  SmSF in Fig.~\ref{fig:opt_cond}d). In SmSF, according to the spectral function shown in Fig.~\ref{fig:bands}c), we have an absorption edge formed by S-3$p$ to Sm-4$f$ optical transitions. These transitions involving 4$f$ states contribute equally to the [100] and [001] optical conductivity, as can be seen at the onset of conductance in Fig.~\ref{fig:opt_cond}d).  
The opposite situation one can observe in GdSF with its pure $p$-$d$ optical gap. As can be seen in Fig.~\ref{fig:opt_cond}e), the [001] contribution to the optical conductivity of GdSF remains very weak up to \unit[5]{eV}, due to the absence of Gd-4$f$ states in this energy range. 

These findings suggest that optical transitions involving the flat and uniform 4$f$ bands do not show any pronounced dependence on the polarization of the incident light. Optical transitions to the highly dispersive 5$d$ band, present in the unoccupied part of the $Ln$SF spectral function, instead are much more susceptible to the polarization and contribute mainly to the [100] optical conductivity. This opens the way for determining the character of the optical absorption edge from measurements of the optical conductivity. Given that high-quality single-crystals are available, the character of optical transitions ($p$-$d$, $p$-$f$) can be identified by measuring the optical conductivity depending on the polarization of the incident light.  

In conclusion, we have presented an \textit{ab initio} approach for calculating the electronic structure and optical response in rare-earth based semiconductors. This approach, combining a dynamical mean-field theory treatment of strong local correlations with a  semi-local exchange-correlation potential, reveals
a non-trivial evolution of the absorption edge along the $Ln$SF series, as evaluated from the one-electron $\vk$-resolved many-body spectral functions and from the polarization-resolved optical conductivities. The nature of the absorption edge is shown to be  reflected in the polarization dependence of
the conductivity, which is dramatically different in the case of
an optical gap of $p \to f$ character (as in SmSF) as compared to the $p \to d$ one (GdSF). Interestingly, we identify two cases (PrSF and NdSF), in which the  $p \to d$ and $p\to f$ transitions are  both simultaneously  in play, resulting in an intermediate behavior.
The present computational methodology and polarization-based analysis of the optical conductivity can be applied to a wide range of rare-earth based materials with a $p$-$d$ band-gap and localized 4$f$ states to identify the nature of the optical gap and to search for new  materials with
potentially useful optical properties, 
including -- but not limited to --
new environmentally-friendly inorganic pigments.

{\em Acknowledgments.}
This work was supported by the Austrian Science Fund (FWF), Schr\"odinger fellowship J-4267, the European Research Council (project 617196 CORRELMAT) and IDRIS-GENCI Orsay under project t2020091393.
We thank the computer team at CPHT for support.

%\bibliography{main}

\end{document}